\documentclass[%
 aip,
 jcp,%
 amsmath,amssymb,
 reprint,%
author-year,%
author-numerical,%
]{revtex4-1}

\usepackage{graphicx}% Include figure files
\usepackage{dcolumn}% Align table columns on decimal point
\usepackage{bm}% bold math
\usepackage{url}

\begin{document}
%\preprint{AIP/123-QED}
\title[\textit{A. Metere, et al.} - Formation of a New Archetypal MOF from a Simple Monatomic Liquid]{Formation of a New Archetypal Metal-Organic Framework from a Simple Monatomic Liquid}

\author{Alfredo Metere}
\email{alfredo.metere@mmk.su.se}

\author{Peter Oleynikov}
%\email{peter.oleynikov@mmk.su.se}

\affiliation{Department of Materials and Environmental Chemistry, Stockholm University, S-106 91, Stockholm, Sweden}%

\author{Mikhail Dzugutov}
%\email{mik@pdc.kth.se}
\affiliation{Department of Mathematics, Royal Institute of Technology,
  S-100 44 Stockholm, Sweden}

\author{Michael O'Keeffe}
%\email{mokeeffe@asu.edu}
\affiliation{Department of Chemistry and Biochemistry, Arizona State 
University, Tempe, Arizona 85287, United States}

\date{\today}

\begin{abstract}
We report a molecular-dynamics simulation of a single-component
  system of particles interacting via a spherically symmetric
  potential that is found to form, upon cooling from a liquid state,
  a low-density porous crystalline phase. Its structure analysis
  demonstrates that the crystal can be described by a net with a
  topology that belongs to the class of topologies characteristic of
  the metal-organic frameworks (MOFs). The observed net is new, and it
  is now included in the Reticular Chemistry Structure Resource
  database (RCSR). The observation that a net topology characteristic
  of MOF crystals, which are known to be formed by a
  coordination-driven self-assembly process, can be reproduced by a
  thermodynamically stable configuration of a simple single-component
  system of particles opens a possibility of using these models in
  studies of MOF nets. It also indicates that structures with MOF
  topology, as well as other low-density porous crystalline structures
  can possibly be produced in colloidal systems of spherical
  particles, with an appropriate tuning of interparticle interaction.
\end{abstract}

\maketitle

\section{Introduction}
Thermodynamic behavior, atomic structure, properties and formation
mechanisms of low-density condensed-matter phases remain a generally
unexplored area. Due to the special properties of low-density porous
phases, this area is in the focus of materials chemistry, materials
science and statistical physics of condensed matter. Besides
conceptual interest related to the peculiarities of the phase
behavior of condensed-matter systems at low densities, this area
presently attracts a great deal of research activity because of
potentially vast scope of technological applications of porous
crystalline materials.  The number of newly discovered structures of such kind, which are reported every year, grows rapidly. Description,
classification and design of these structurally complex materials can arguably be regarded as the last frontier of crystallography.

The current explosive growth in the number of newly created extended
porous crystalline structures is a result of the development of
reticular synthesis, a crystal engineering approach whereby
pre-designed stable building blocks\cite{SBUs} are assembled into a
preliminarily conceived periodic network. In particular, this
development greatly advanced the synthesis of metal-organic frameworks
(MOFs) that attract special interest because of the technological
importance of these materials\cite{okeeffe1,okeeffe2}. MOFs
represent coordination networks with organic ligands containing pores
that can be used e.g. for hydrogen storage, or gas separation.

The extensive research efforts aimed at the design of new MOFs raise
the problem of their modelling and systematic structural
characterization. The general purpose of mathematical modelling of
real systems consists in reduction of the observed
complexity. Reproduction of a property of an investigated system by a
simplified model allows to discern its essential aspects; in this way,
it makes it possible to understand the observed phenomenon by putting
it in a general context which can be formalized in a respective
classification.  Classical crystallography is an example of such an
approach: it provides a unifying description of the periodic atomic
structures in terms of a general crystallographic classification,
thereby reducing their compositional complexity to simplified set of
archetypal geometrical model configurations.

Recently, a conceptually new approach to the description of extended
periodic structures has been developed. It exploits the idea of
presenting these structures in terms of topology of infinite periodic
graphs (nets) which are assumed to represent the networks of
bonds\cite{SBUs} connecting the elementary structural blocks. This is
an alternative to the traditional crystallographic method of structure
characterization whereby the crystalline structures are considered in
terms of the geometry of configurations of points.  The development of
this new concept of structure description has been prompted by the
advent of reticular chemistry. Using the structure description in
terms of the topology of bonds, hypothetical MOFs can be redesigned
from the existing structures by recombining the building blocks
according to the conceived net topology\cite{okeeffe4, WILL}.

A basic question arising in the evaluation of a hypothetical MOF model
produced as a decoration of a net design concerns its mechanical and
thermodynamic stability. This implies that the atomic configurations
designed in that way are expected to represent both potential-energy
minimum, and free-energy minimum. When investigating this question,
the use of a direct particle simulation of a MOF model producing a
realistic atomic configuration is indispensable \cite{CAB}. However,
such direct simulations of MOF structures are critically constrained
by the extreme complexity of the actual MOF structure. This raises a
question of conceptual interest: how simple could a physical model,
reproducing topology of a MOF net, be, and to what extent can the
complexity of a real atomic structure described by a net be reduced?
In particular, can its compositional complexity be reduced to a
single-component simple system of particles?

These questions are addressed in the molecular-dynamics simulation which we describe here.
It is demonstrated that a single-component system of particles interacting via a spherically symmetric potential can form a thermodynamically stable crystalline structure with a net topology that belongs to a class of nets commonly observed in MOFs.
In this way, the simulation happened to produce a new archetypal MOF with a novel type of net that is now included in the RCSR database. In
the following, we present a detailed description of this simulation
and its results. We also discuss possible use of the general
simple-system approach exploited in the present model for the topology
design and analysis of other low-density porous crystalline
structures, including experiments with colloidal systems.

\section{Model and Simulation}
As we have explained above, the general purpose of the simulation we
report here is to explore the idea of self-assembly formation of
low-density porous crystalline structures in systems of identical
particles using a spherically-symmetric interaction potential. For
such purpose, we conceived the design of an interparticle pair
potential in accordance with the following guidance concepts. 

First, it is obvious that a mechanically stable energy-minimum
low-density configuration of particles must necessarily possess a
reduced number of neighbors in the first coordination shell, as
compared with densely packed structures. In a system with a
spherically-symmetric interaction, this reduction can be achieved by
constraining the radial variations in the first-neighbor distance
which are characteristic of the dense sphere packing. Such a
constraint can be arranged by confining the first neighbors to a
potential energy minimum of sufficiently reduced width. 

Second, the separation between the first neighbor shell and the rest
of the structure defines the configuration's density and
porosity. This is controlled by the long-range repulsive branch of the
pair potential. To produce this structural effect, the first minimum
is supposed to be followed by a sufficiently broad maximum.

Core-softened pair potentials possessing an additional long-range
repulsion have been known to induce non-trivial phase behavior
\cite{FO1,FO2}, isostructural transitions\cite{ST1,ST2}, liquid-liquid
transitions, formation of clusters\cite{DOB}, columnar and lamellar
structures\cite{PAU1,PAU2}. A soft-core form of pair potential has
also been found to produce formation of a glassy state \cite{DZ1} and
a quasicrystal \cite{DZ2}. Isotropic soft-core potentials have also
been used to produce low-coordinated structures \cite{TOR, PRES, ZH}

\begin{figure} %figure 1
\includegraphics[width=6.2cm]{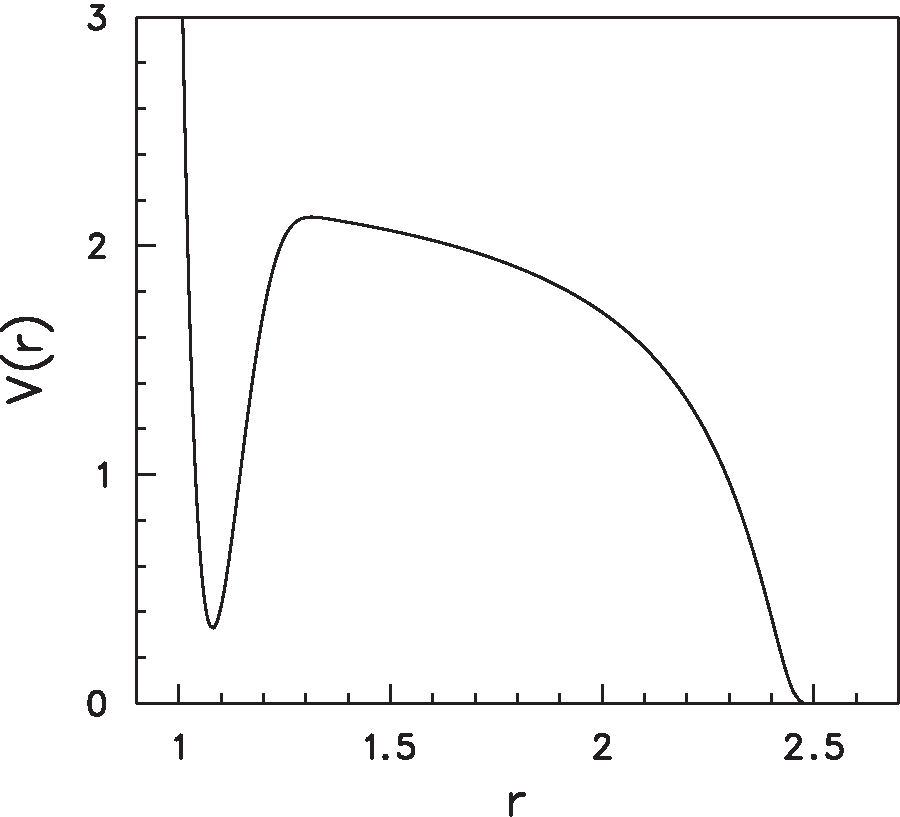}
\caption{Pair potential}
\label{fig1}
\end{figure}

These considerations were incorporated in the design of the
spherically symmetric interparticle potential we exploited in this
study.  It is shown in in Figure \ref{fig1}. The functional form of
the potential energy for two particles separated by the distance $r$
is:

\begin{equation} 
V(r)  = a_1  ( r^{-m} - d) H(r,b_1,c_1) + a_2 H(r,b_2,c_2)  \\ 
\end{equation}
where
\begin{equation} 
H(r,b,c) = \left\{ \begin{array}{ll}
  \exp\left( \frac{b}{r-c} \right) & r < c\\
0 & r \geq c
\end{array}
\right.       
\end{equation}

\begin{figure} %figure 2
\includegraphics[width=8.cm]{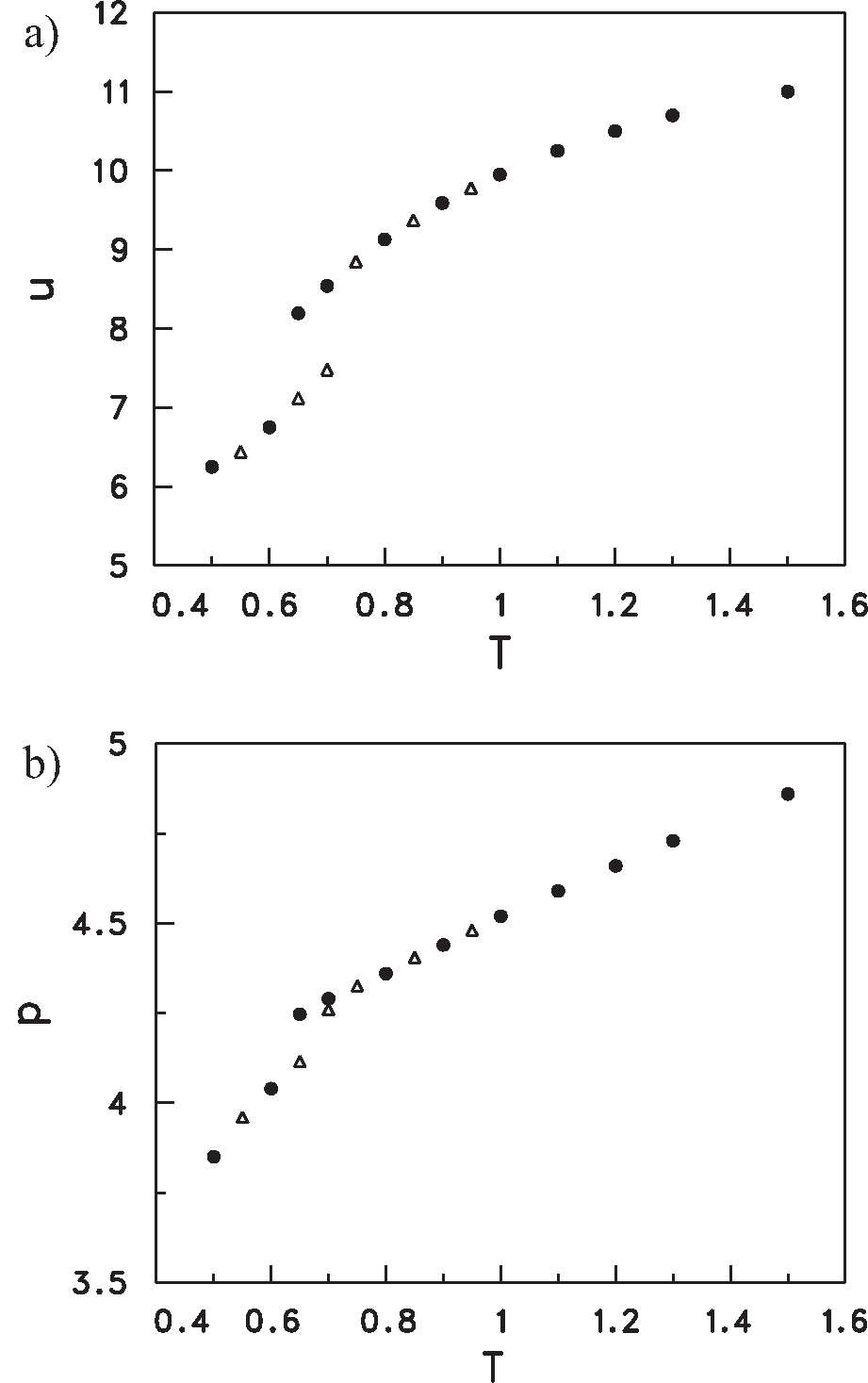}
\caption{Isochoric liquid-solid phase transformation. $a)$ and $b)$,
  respectively, depict energy and pressure variation as a function of
  temperature. Dots and open triangles correspond to cooling and
  heating, respectively.}
\label{fig2}
\end{figure}

The values of the parameters are presented in Table \ref{table1}. The first term
of this functional form describes the short-range repulsion branch of
the potential, and its minimum, whereas the second term is responsible
for the long-range repulsion part. We remark that all the simulation
results we report here are expressed in terms of the reduced units
that were used in the definition of the potential. We also note that
the short-range repulsion branch of the potential, and the position of
its first minimum closely approximate those in the Lennard-Jones (LJ)
potential \cite{hansen}, which makes it possible to compare directly
the reduced number densities, and other thermodynamic quantities of
the two systems.
\begin{table}
\begin{tabular}{cccccccc}
 
\hline 
\hline 
m & $a_1$  & $b_1$ & $c_1$ & $a_2$ & $b_2$ & $c_2$ & $d$ \\
\hline 
\hline 
12 \space \space & 265.85 \space \space & 1.5 \space \space & 1.45 \space 
\space
& 2.5 \space \space &
0.19 \space \space & 2.5 \space \space & 0.8 \\
\hline \\
\end{tabular}
\caption{Values of the  parameters for the pair potential.} 
\label{table1}
\end{table}

We note that this pair potential represents a modification of an
earlier reported one designed using the same functional form with a
shorter repulsion range which was found to produce a Smectic-B
crystal\cite{ME}.

In this study, we employed molecular-dynamics simulation to
investigate thermodynamic behavior of a system of identical particles
interacting via the described pair potential at low density.  The
system, comprising 16384 particles, was confined to a cubic box with
periodic boundary condition. The simulation was carried out at
constant number density $\rho=0.3$, in reduced units. We note that
this density is very low if estimated in comparison with the LJ system
phase behavior: the LJ triple-point density occurs at $\rho=0.84$
\cite{hansen}.

At the beginning of the simulation, the system has been equilibrated
in its thermodynamically stable isotropic liquid state at sufficiently
high temperature at the number density $\rho=0.3$.
The liquid was then subjected to isochorical cooling.
This was performed in a stepwise manner: the system was comprehensively equilibrated at fixed temperature after each cooling step.
As the liquid has been cooled below $T=0.65$, a discontinuous change in the thermodynamic parameters was detected, which can be seen in Fig. \ref{fig2}.
The thermodynamic singularity was found to be accompanied by a sharp drop in the rate of self-diffusion.
These observations represent an apparent signature of a first-order phase transition to a solid phase.
The conclusion was further confirmed by a significant hysteresis that was produced when heating the low-temperature phase.
A non-trivial character of the low-temperature phase was indicated by an
anomalously long time required for its equilibration which amounted to
several billions of time-steps.

\section{Results}

\subsection{Structure}
We now turn to characterization of the simulated structure, and try to 
understand its relationship with the pair potential.
In order to facilitate the structure characterization, we analyzed an energy-minimum configuration.
For that purpose, the steepest-descent minimization procedure was applied to a system's instantaneous configuration thereby reducing the system to its nearest potential-energy minimum.
In this way, the configuration was rid of any thermally induced perturbations.

\begin{figure}[h] %figure 3
\includegraphics[width=8cm]{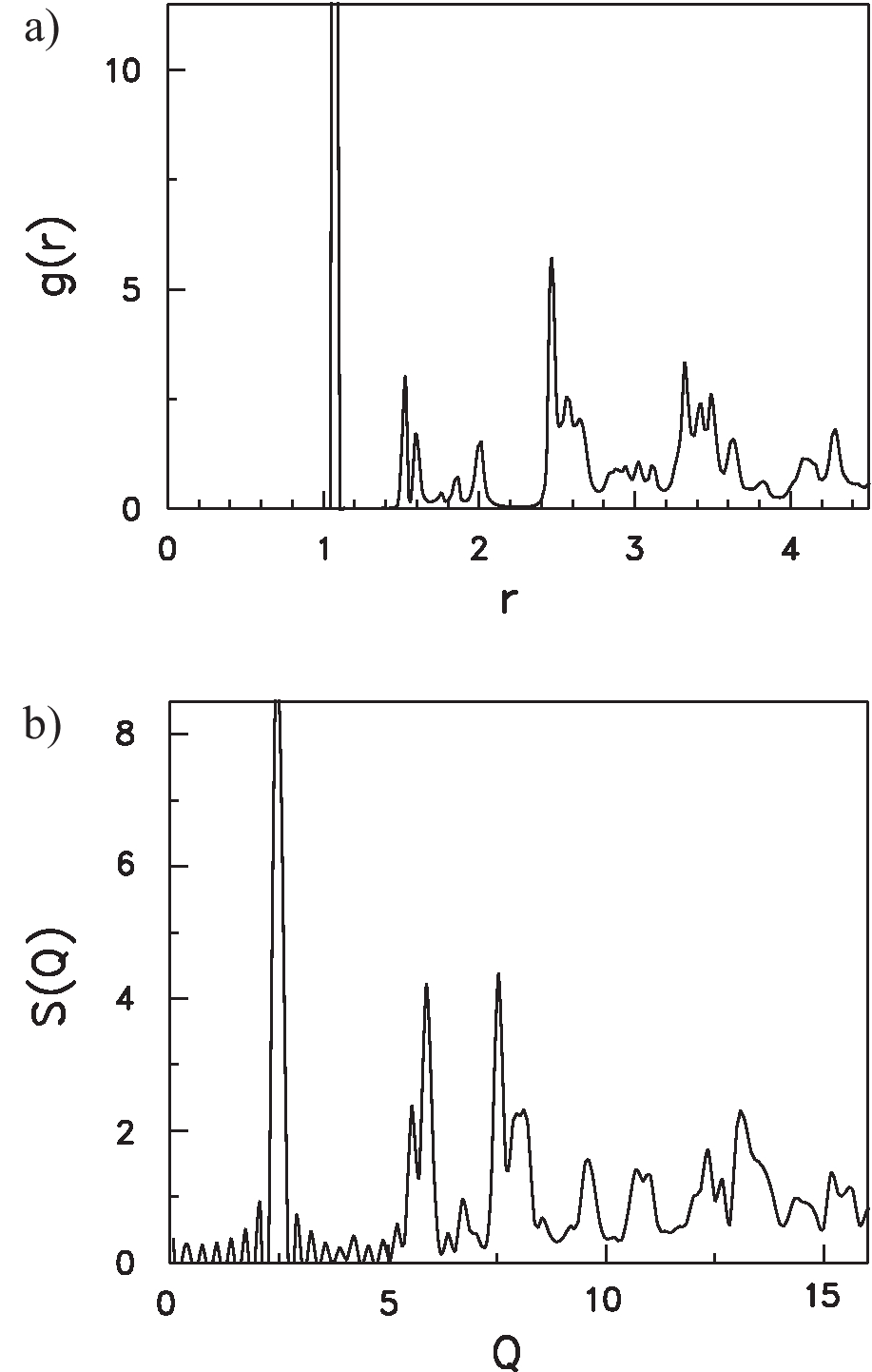}
\caption{ $a)$ Radial distribution function. $b)$ Spherically
  averaged structure factor.}
\label{fig3}
\end{figure}

\begin{figure} %figure 4
\includegraphics[width=7.15cm]{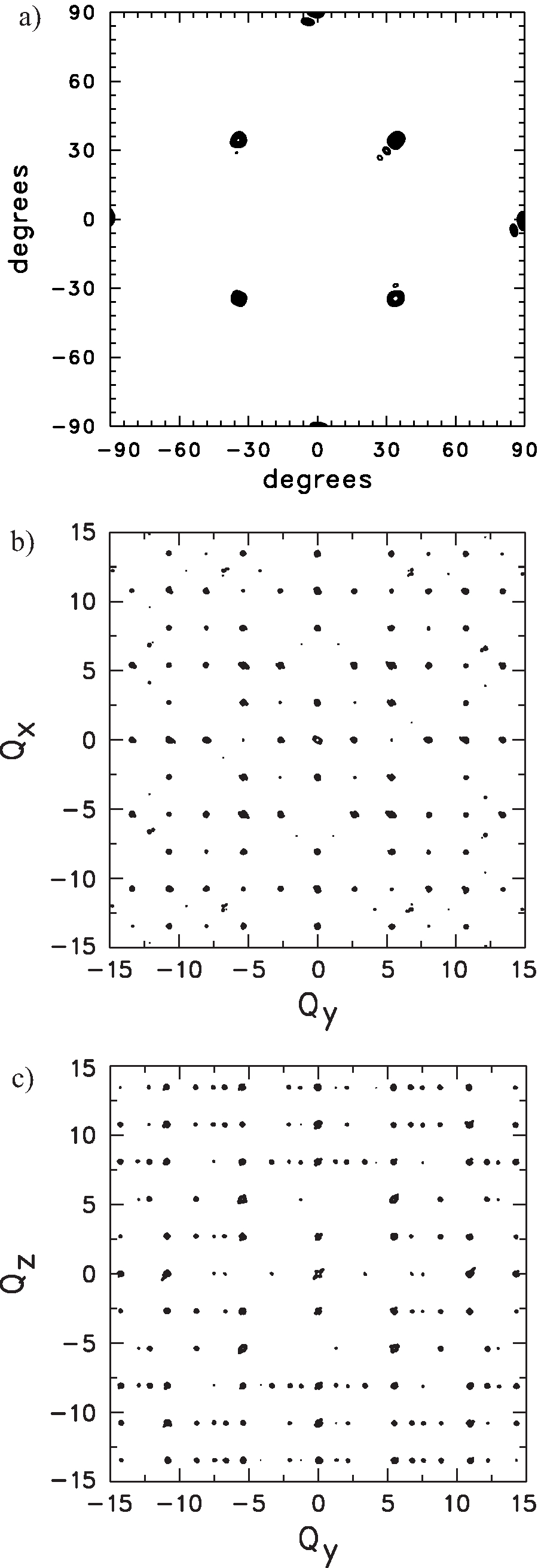}
\caption{ The isointensity plots of the structure factor $S(\bf
  Q)$. $a)$ Diffraction intensity in the reciprocal-space sphere
  of the radius $Q=2.5$ which corresponds to the position of the sharp
  pre-peak of $S(Q)$ in Fig. \ref{fig2}, as viewed from the axial
  direction. $b)$ In the reciprocal-space plane orthogonal to
  the axis, $Q_z=0$. $c)$ In the axially oriented plane,
  $Q_y=0$. $Q_z$ denotes the axial dimension, and $Q_y$ corresponds to
  a translational symmetry vector, orthogonal to the axis.}
\label{fig4}
\end{figure}
The structure was  first inspected  in terms of one-dimensional correlation functions, the radial distribution function, $g(r)$, and its
reciprocal-space counterpart, the spherically-averaged structure factor $S(Q)$ which are related as: 
\begin{equation}
S(Q) = 1 + 4 \pi \rho \int_{0}^{\infty} [g(r)-1] \frac{\sin(Qr)}{Qr} r^2 dr
\end{equation}

These correlation functions are presented in Fig. \ref{fig3}. An apparent
anomaly featured by the calculated $g(r)$ is a big gap between the
first and the second meaningful maximum; the latter occurs at the
distance of about $r=2.5$. The big separation distance to the second
shell of neighbors, anomalous for a dense sphere packing, can be
associated with the characteristic void size in a porous
structure. Accordingly, $S(Q)$ features a singularly sharp peak at the
$Q$-value corresponding to the indicated distance, which can
apparently be regarded as a characteristic length-scale of this porous
configuration.

Further insight into the origin of these structural peculiarities can
be obtained from the analysis of the Fourier-space distribution of the
structure factor  $S({\bf Q})$. The latter is calculated as:

\begin{equation}
S({\bf Q}) =
\frac{1}{N}\langle \rho({\bf Q})\rho(-{\bf Q})\rangle
\end{equation}
where $\rho({\bf Q})$ is a Fourier-component of the system's number
density:
\begin{equation}
\rho({\bf Q}) =  \sum_{j=1}^{N}\exp (i{\bf Q r}_j) 
\end{equation} 
${\bf r}_i$ being the positions of the system's particles, and $N$ is
the number of particles in the system. Notice that $S({\bf Q})$
represents the signal intensity as measured in diffraction
experiments.

As a first step, we calculated the diffraction intensity on the
$Q$-space sphere of diameter corresponding to the position of the
sharp pre-peak of the spherically averaged $S(Q)$, see
Fig. \ref{fig3}. This result is shown in Fig. \ref{fig4}(a). 
It demonstrates that the pre-peak decomposes into a
set of well-defined diffraction maxima which form a regular four-fold
pattern. This made it possible to determine the global symmetry of
the configuration, which is found to be characterized by a single
four-fold axis. The axis orientation having been determined, we
calculated $S({\bf Q})$ within two characteristic $Q$-space planes:
$Q_z=0$ and $Q_y=0$, $Q_z$ being the axis coordinate, and $Q_y$
coordinate corresponding to a translational symmetry vector orthogonal
to the axis. These two diffraction planes are also shown in
Fig. \ref{fig4}(b, c).

The pattern of diffraction maxima presented in Fig. \ref{fig4} is
visibly dominated by the characteristic spacing $\Delta Q=2.5$ that
has been detected as the position of the first sharp peak of $S(Q)$ in
Fig. \ref{fig3}. This feature we attributed to the presence of voids
with characteristic spacing of about $2.5$ atomic diameters. It can
thus be concluded that these voids represent the dominating element of
the investigated structure, forming a periodic pattern with single
four-fold axis.

\subsection{Description of the real-space configuration of our simulated
structure}
In this section we present a real-space description of the simulated
phase structure. We discern its unit cell and demonstrate the
structure's porosity which was conjectured in the previous section
from the diffraction pattern.

The particle configuration, representing the simulated crystal, is
confined to a cubic simulation box with periodic boundary
conditions. We note that the crystal is misaligned with respect
to the coordinate axes of the confining box. 

% Unit cell, RDF, pores

%%% UNIT CELL

\begin{figure}
\includegraphics[width=8cm]{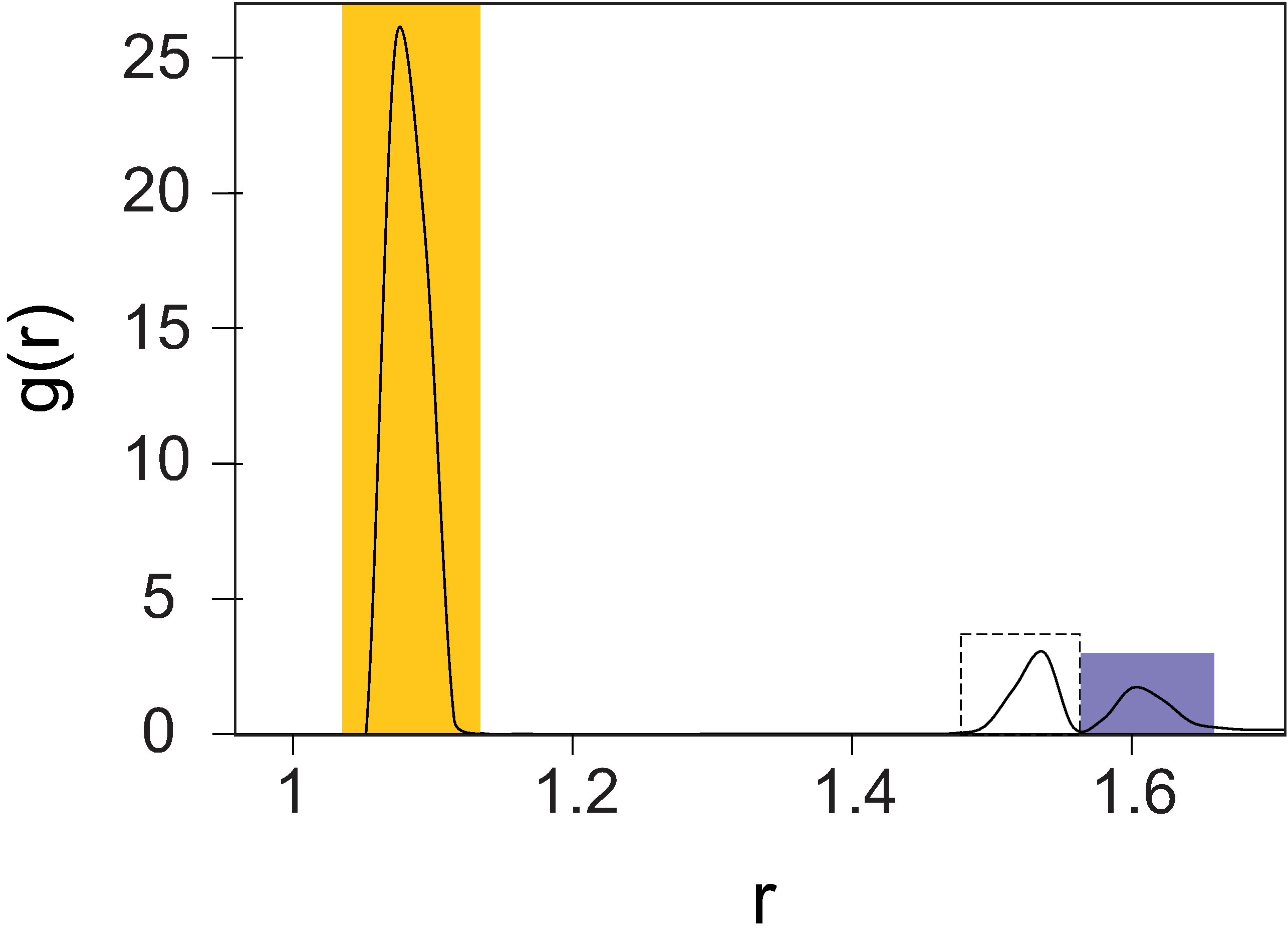}
\caption{Portion of the radial distribution function of the simulated
  crystal configuration. Highlighted are the distance ranges
  corresponding to the peaks which are used as bonds defining the unit
  cell Fig. \ref{fig6}. Yellow and purple, respectively, depict the
  peaks positioned at $r_{1}\approx1.08$ and $r_{2}\approx1.63$.}
\label{fig5}
\end{figure}

As the first step of the real-space structure characterization we
describe the correlations of the particle positions in terms of the
radial distribution function (RDF). The short-range part of this
function is presented in Fig. \ref{fig5}. The distribution reveals a
set of peaks that represent the characteristic distances between the
nearest neighbors. Two of these peaks, highlighted in Fig. \ref{fig5}
by colors, represent two ranges of interparticle distances which are
defined as the basic bonds forming the structure. The simulated
crystal is then described as a network of these bonds.

\begin{figure}
\centering
\includegraphics[width=6cm]{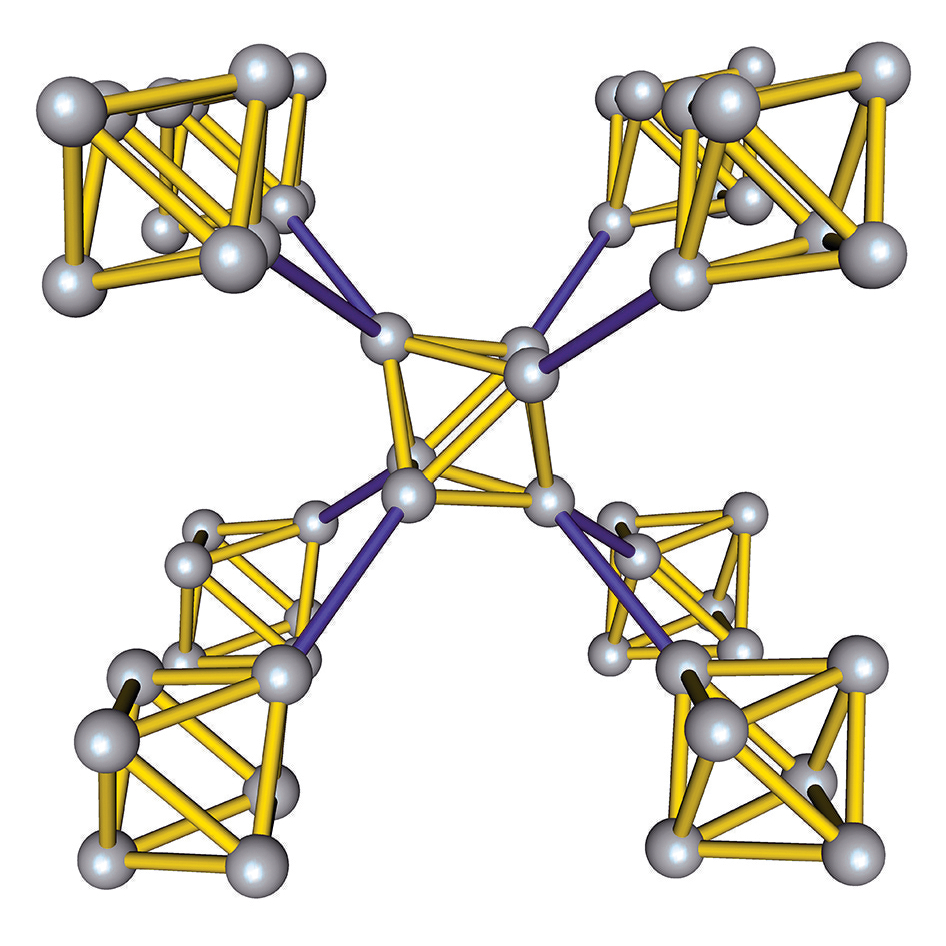}
\caption{The unit cell of the simulated crystal configuration. The
  yellow and purple bonds correspond to the distance ranges higlighted
by similar colors in Fig. \ref{fig5}}
\label{fig6}
\end{figure}

Next, we consider thus defined network of bonds to discern the unit
cell the periodic repetition of which in three dimensions forms the
simulated crystal structure. The unit cell configuration is shown in
Fig. \ref{fig6}.  Its bonds are colored in the same manner as that used
to highlight the respective peaks of the radial distribution function
shown in Fig. \ref{fig5}.

% Begin Figure -------------------------------------------------
% End figure ---------------------------------------------------
The unit cell parameters and the positions of the asymmetric unit
vertices, scaled by the size of the unit cell are reported in Table
\ref{table2}. These were estimated in terms of the interparticle
distances corresponding to the maxima of the highlighted peaks in
Fig. \ref{fig5}.  The unit cell is tetragonal and has $P4_2/mnm$
symmetry, which explains why there are only two vertices in the
asymmetric unit.

\begin{table}
\begin{tabular}{cccc}
\hline
\hline
Space group & $a=b$ & $c$ & $\alpha=\beta=\gamma$\\
\hline
\hline
\space $P4_{2}/mnm$ \space & 3.600 \space & 3.200 \space & \space 90$^\circ$
\space \\
\hline \\
\end{tabular}

\begin{tabular}{cccc}
\hline
\hline
Vertex & x/a & y/b & z/c\\
\hline
\hline
\space $V_1$ \space & \space 0.39342 \space & \space 0.39342 \space & \space 0.33077 \space \\
\space $V_2$ \space & \space 0.14994 \space & \space 0.14994 \space & \space 0       \space \\
\hline \\
\end{tabular}
\caption{Reconstructed unit cell parameters and list of the vertices in the
asymmetric unit, scaled by the unit-cell size.}
\label{table2}
\end{table}

Connecting particles with bonds of $r_1\approx1.08$ length will
generate a regular octahedron in the middle of the unit cell.  Such
bonds correspond to the first peak position in the radial distribution
function which is related with the 3D periodic pattern.  Bonds of the
length $r_2\approx1.63$ will connect the central octahedron with its
symmetry-related replicas located at the corners of the unit cell.

The four basal particles of each octahedron are 5-coordinated, while
the two apical particles are 6-coordinated.  A graphical
representation of the unit cell with the particles connected according
to the described bonding scheme is shown in Fig. \ref{fig6}.\

The peak region presented in the radial distribution function (see
Fig. \ref{fig5}) as the dashed region corresponds to the interparticle
distances distribution representing the solid (volume) diagonals of the
octahedra.  However, we did not take this peak region into the account
when constructing the bond network because the resulting bonds would
intersect with each other in the centers of the octahedra.

\begin{figure}
\includegraphics[width=8.0cm]{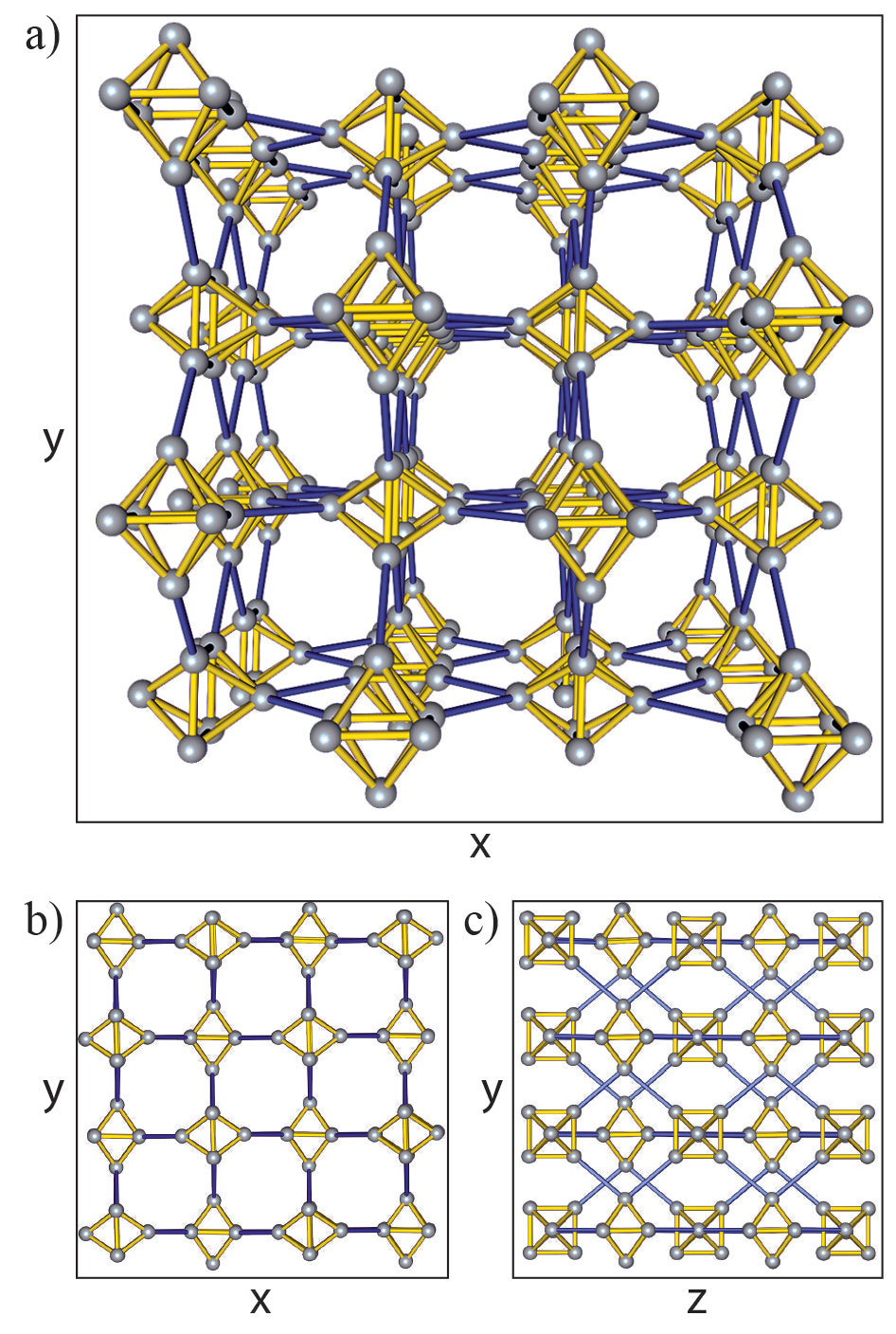}
\caption{A fragment of the simulate crystal configuration produced
  using the same bonding method as used for the unit cell plot in Fig.
  \ref{fig6}, using the same respective bond colors. $a)$
  Perspective projection demonstrating how the octahedra are
  interconnected to form the pores. $b)$ and $c)$: orthogonal
  projections.}
\label{fig7}
\end{figure}

A fragment of the simulated crystal configuration presented as a
network of bonds is shown in Fig. \ref{fig7}. By considering the
sequences of adjacent unit cells it is possible to notice how their
self-repetition along the unit cell axes generates a mono-dimensional
channel system which develops along the $Z$-axis. The pore diameter
is estimated as $\phi \approx 2.5$, which is comparable with the size
of the unit cell.  We also remark that the pore diameter is
consistent, as anticipated, with the ultimate distance of the second
(longer-range) repulsion part of the pair potential.

It is not directly possible to assess the class of real materials
which our simulated crystal structure belongs to. This is because the
definitions of micro-, meso- and macro-porosity are related to the
size of the pore diameter expressed in explicit units rather than
reduced units.\ Proceeding by exclusion we can assert that our
structure does not belong to any known zeolitic, Covalent Organic
Frameworks (COFs) and Zeolitic Imidazolate Frameworks (ZIFs) types,
because of the coordination numbers of its particles.\ The periodically
ordered pore walls indicate that our structure is not to be classified
as crystalline mesoporous silica either.  Excluded this way the other
candidates, we conclude that our structure is of a new archetypal MOF.

\begin{figure}
\includegraphics[width=8cm]{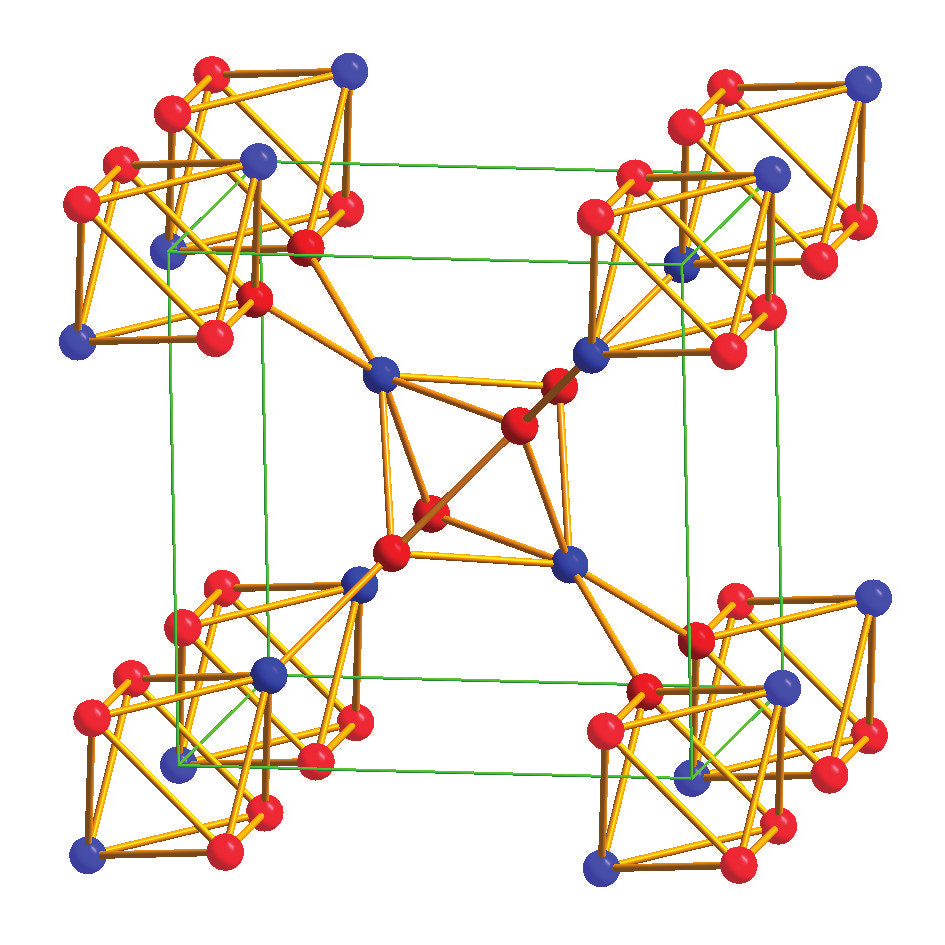}
\caption{The unit-cell configuration for the 'alm' net included in
  RCSR classification \cite{RCSR}, its parameters presented in Table
  \ref{table3} (see explanations in the text)}
\label{fig8}
\end{figure}

\subsection{The  net classification}
We now turn to the topological classification of the bond network of
our simulated particle configuration.  Our structure exhibits a novel
topology (net) that has been included in the Reticular Chemistry
Structure Resource (RCSR) database\cite{RCSR} where it is named
'$\bf{alm}$'. That crystal net produced using the method of bonding
that we described above is shown in Fig. \ref{fig8}.  The connectivity
of the nodes in this unit cell is exactly the same as was observed in
the the unit cell of the simulated structure shown in Fig. \ref{fig6}.
The only difference between the two unit cells configurations is that
in the one shown in Fig. \ref{fig8} all the bond length values have
been we normalized to 1.  The '$\bf{alm}$'-net unit cell parameters
and the asymmetric unit vertices are presented in Table \ref{table3}.

\begin{table} [h]
\begin{tabular}{cccc}
\hline
\hline
Space group & $a=b$ & $c$ & $\alpha=\beta=\gamma$\\
\hline
\hline
\space $P4_{2}/mnm$ \space & 2.7818 \space & 2.3002 \space & \space 90$^\circ$
\space \\
\hline \\
\end{tabular}

\begin{tabular}{cccc}
\hline
\hline
Vertex & x/a & y/b & z/c\\
\hline
\hline
\space $V_1$ \space & \space 0.3729 \space & \space 0.3729 \space & \space
0.2826 \space \\
\space $V_2$ \space & \space 0.1797 \space & \space 0.1797 \space & \space 0
\space \\
\hline \\
\end{tabular}
\caption{Unit cell parameters and vertices of the 'alm' net included
  in the RCSR classification\cite{RCSR}.}
\label{table3}
\end{table}

\section{Conclusions}
The simulation results we report here demonstrate that a
thermodynamically stable, low-density, microporous, crystalline phase,
exhibiting a kind of topology characteristic of the MOF nets, can be
formed by a single-component system of particles with a
spherically-symmetric interaction.

A more general conclusion of this study is that the form of pair
potential we explored here can be further exploited to simulate other
novel extended low-density porous crystalline structures, including
those with MOF topologies. This approach can possibly be regarded as
complementary to the mathematical efforts, in exploring the topological
variety of possible reticular chemistry structures.

Another perspective application of this simulation is that it can be
used as a guidance for creating similar structures in colloidal
systems of spherical particles with appropriate tuning interparticle
force field. It has to be mentioned that the main features of the pair
potential we exploited in this simulation are consistent with the
classical theory for colloidal interactions by Deryagin, Landau,
Verwey and Overbeek (DLVO) \cite{DL,DL1,DL2,MAL}, amended with hard-core
repulsion or steric repulsion close to the contact.

\section{Acknowledgments}
The authors thank Sten Sarman for illuminating discussions. The
simulations were performed using the resources provided by the Swedish
National Infrastructure for Computing (SNIC) at PDC Centre for High
Performance Computing (PDC-HPC).

\end{document}